\documentclass[sigconf, screen]{acmart}

\usepackage{graphicx} %
\usepackage{natbib}  %
\usepackage{caption} %
\usepackage{algorithm}
\usepackage{listings}
\usepackage{algorithmic}
\usepackage{amsmath}
\usepackage{booktabs}
\usepackage{multirow}
\usepackage[outdir=./]{epstopdf}
\usepackage{enumitem}
\usepackage{caption}
\usepackage{subcaption}
\usepackage{newfloat}
\usepackage{float}
\usepackage{balance}
\usepackage{svg} % svg
\usepackage[normalem]{ulem}
\usepackage{threeparttable}
\usepackage{multirow} % 用于 multirow
\usepackage{xcolor}
\definecolor{darkred}{rgb}{0.5,0,0}
\definecolor{darkblue}{rgb}{0,0,0.5}
\definecolor{darkgreen}{rgb}{84,130,53}

\AtBeginDocument{%
  }

\begin{document}

%%
%% The "title" command has an optional parameter,
%% allowing the author to define a "short title" to be used in page headers.
% \title{LLM-assisted Explicit and Implicit Multi-interest Learning Framework for Sequential Recommendation}
% \title{Alignment of LLM Semantics and User Behavior for  Multi-interest Modeling}
% \title{A Bi-level Multi-interest Learning Framework Combining LLM and Sequential Recommendation Model}
% \title{Rethinking Semantic IDs for Generative Recommendation: \\ From Learned Codebooks to LLM-aligned Token Sets}
\title{VaLiDRec: Variable-Length LLM-Aligned Semantic IDs for Generative Recommendation}
% \acmSubmissionID{688}
% \shorttitle{LLM-assisted Explicit and Implicit Multi-interest Representation Learning Framework}

%%
%% The "author" command and its associated commands are used to define
%% the authors and their affiliations.
%% Of note is the shared affiliation of the first two authors, and the
%% "authornote" and "authornotemark" commands
%% used to denote shared contribution to the research.
% \author{Ben Trovato}
% \authornote{Both authors contributed equally to this research.}
% \email{trovato@corporation.com}
% \orcid{1234-5678-9012}

% \author{G.K.M. Tobin}
% \authornotemark[1]
% \email{webmaster@marysville-ohio.com}
% \affiliation{%
%   \institution{Institute for Clarity in Documentation}
%   \city{Dublin}
%   \state{Ohio}
%   \country{USA}
% }

\author{Shutong Qiao}
% \orcid{0000-0002-7368-1535}
\affiliation{
  \institution{University of Queensland}
  \city{Brisbane}
  \country{Australia}
}
\email{shutong.qiao@uq.edu.au}

\author{Wei Yuan}
\affiliation{
  \institution{University of Queensland}
  \city{Brisbane}
  \country{Australia}
}
\email{w.yuan@uq.edu.au}

\author{Tong Chen}
\affiliation{
  \institution{University of Queensland}
  \city{Brisbane}
  \country{Australia}
}
\email{tong.chen@uq.edu.au}

\author{Hao Wang}
\affiliation{
  \institution{Computer Network Information Center, Chinese Academy of Sciences}
  \city{Beijing}
  \country{China}
}
\email{cashenry@126.com}

\author{Quoc Viet Hung Nguyen}
\affiliation{
    \institution{Griffith University}
    \city{Gold Coast}
    \country{Australia}
}
\email{henry.nguyen@griffith.edu.au}

\author{Hongzhi Yin}
\authornote{Corresponding author.}
\affiliation{
    \institution{University of Queensland}
    \city{Brisbane}
    \country{Australia}
}
\email{h.yin1@uq.edu.au}

\renewcommand{\shortauthors}{Shutong Qiao et al.}
%%
%% The abstract is a short summary of the work to be presented in the
%% article.
% \begin{abstract}
% The code repository is publicly available at the anonymous link: \url{https://anonymous.4open.science/r/EIMF}.

% \end{abstract}
\begin{abstract}
Generative recommendation commonly represents items using fixed-length semantic identifiers (SIDs) constructed through clustering and quantization. However, these artificial codes may overcompress item semantics, remain misaligned with pretrained LLM vocabularies, and require costly autoregressive decoding. 
In light of this, we propose VaLiDRec, a generative recommendation framework based on variable-length, LLM-aligned semantic identifiers. VaLiDRec constructs SIDs directly from informative native LLM vocabulary tokens via token importance estimation, semantic-quality-aware pruning, and collision-aware refinement, allowing identifier lengths to adapt to item semantic complexity. To model user preferences, VaLiDRec incorporates graph-aware soft prompts and reformulates recommendation as token-set prediction with token-level item scoring, eliminating autoregressive SID generation and beam search.
Experiments on four real-world datasets show that VaLiDRec consistently outperforms strong sequential and generative recommendation baselines across all evaluation metrics. It further achieves superior zero-shot item cold-start performance and 87.49$\times$ faster inference than LC-Rec. These results demonstrate that LLM-native variable-length semantic identifiers provide a more expressive and efficient paradigm for generative recommendation.

%Generative recommendation commonly represents items with semantic identifiers (SIDs). Existing methods typically construct SIDs through clustering or quantization, mapping items into fixed-length latent codes. This process may over-compress information-rich items, discard fine-grained textual semantics, and make semantic retention difficult to assess. Moreover, the resulting artificial codes are not naturally aligned with the vocabulary of pretrained large language models (LLMs), requiring additional adaptation for recommendation. We propose \textbf{VaLiDRec}, a variable-length LLM-aligned SID framework for generative recommendation. VaLiDRec constructs SIDs directly from informative native LLM vocabulary tokens through token importance scoring and semantic-quality-aware pruning, while collision-aware expansion and suffix injection ensure item-level uniqueness. To model user preferences, VaLiDRec aggregates unordered SID tokens into item-level representations and encodes item relations and behavioral transitions with graph-aware soft prompts. It further formulates recommendations as token-set prediction with token-level item scoring, avoiding autoregressive SID generation and beam search. Experiments on four real-world datasets show that VaLiDRec consistently outperforms strong sequential and generative baselines on all evaluation metrics. It also achieves superior zero-shot item cold-start performance and a \(87.49\times\) inference speedup over LC-Rec. An anonymized implementation is available at \url{https://anonymous.4open.science/r/XXXX}.%
\end{abstract}

%%
%% The code below is generated by the tool at http://dl.acm.org/ccs.cfm.
%% Please copy and paste the code instead of the example below.
%%

\begin{CCSXML}
<ccs2012>
<concept>
<concept_id>10002951.10003317.10003331.10003271</concept_id>
<concept_desc>Information systems~Recommender Systems</concept_desc>
<concept_significance>500</concept_significance>
</concept>
</ccs2012>
\end{CCSXML}

\ccsdesc[500]{Information systems~Recommender systems}

% \begin{CCSXML}
% <ccs2012>
% <concept>
% <concept_id>10002951.10003317.10003331.10003271</concept_id>
% <concept_desc>Information systems~Recommender Systems</concept_desc>
% <concept_significance>500</concept_significance>
% </concept>
% </ccs2012>
% \end{CCSXML}

% \ccsdesc[500]{Information systems~Recommender systems}

%%
%% Keywords. The author(s) should pick words that accurately describe
%% the work being presented. Separate the keywords with commas.
\keywords{Recommender System; Generative Recommendation; Sequential Recommendation; Large Language Model}
%% A "teaser" image appears between the author and affiliation
%% information and the body of the document, and typically spans the
%% page.

% \received{20 February 2007}
% \received[revised]{12 March 2009}
% \received[accepted]{5 June 2009}

%%
%% This command processes the author and affiliation and title
%% information and builds the first part of the formatted document.
\maketitle

\section{Introduction}
Generative recommendation (GR) formulates recommendation as a conditional generation problem, where the target item identifier or textual representation is generated directly from user interaction histories, contextual information, and natural language prompts~\cite{geng2022recommendation, wang2023generative, rajput2023recommender, wang2025generative}. Unlike conventional discriminative recommenders that rank candidate items using matching functions, GR integrates user preference modeling, item semantic understanding, and candidate generation within a unified framework. Leveraging the strong language understanding and semantic modeling capabilities of large language models (LLMs)~\cite{brown2020language,grattafiori2024llama,bao2023large}, recent GR methods can directly exploit rich item metadata and pretrained semantic knowledge, making them particularly promising for sparse, long-tail, and cold-start recommendation scenarios.

Generating appropriate item representations is fundamental to generative recommendation, as the output space must be both semantically expressive and precisely retrievable. Directly generating raw item IDs~\cite{geng2022recommendation, hua2023index,petrov2023generative} preserves one-to-one item mapping but lacks semantic structure, making it difficult to capture semantic relationships among items. In contrast, generating free-form textual descriptions provides richer semantic expressiveness but results in an open-ended output space, where generated content may be ambiguous, hallucinated, or difficult to align with items in the corpus. To balance semantic expressiveness, controllability, and inference efficiency, recent GR methods commonly adopt semantic identifiers (SIDs)~\cite{rajput2023recommender, hou2023learning, lin2025order}, which represent each item as a sequence of discrete semantic tokens and train generative models to predict the target item's SID. These SIDs are typically constructed using clustering or quantization methods, such as RQ-KMeans~\cite{deng2025onerec}, RQ-VAE~\cite{lee2022autoregressive}, and FSQ~\cite{mentzer2024finite}, to encode item similarity into a structured output space. Consequently, SIDs have emerged as an effective interface between generative sequence modeling and large-scale item retrieval.

\begin{figure}[t]
    \centering
    \includegraphics[width=1.0\linewidth]{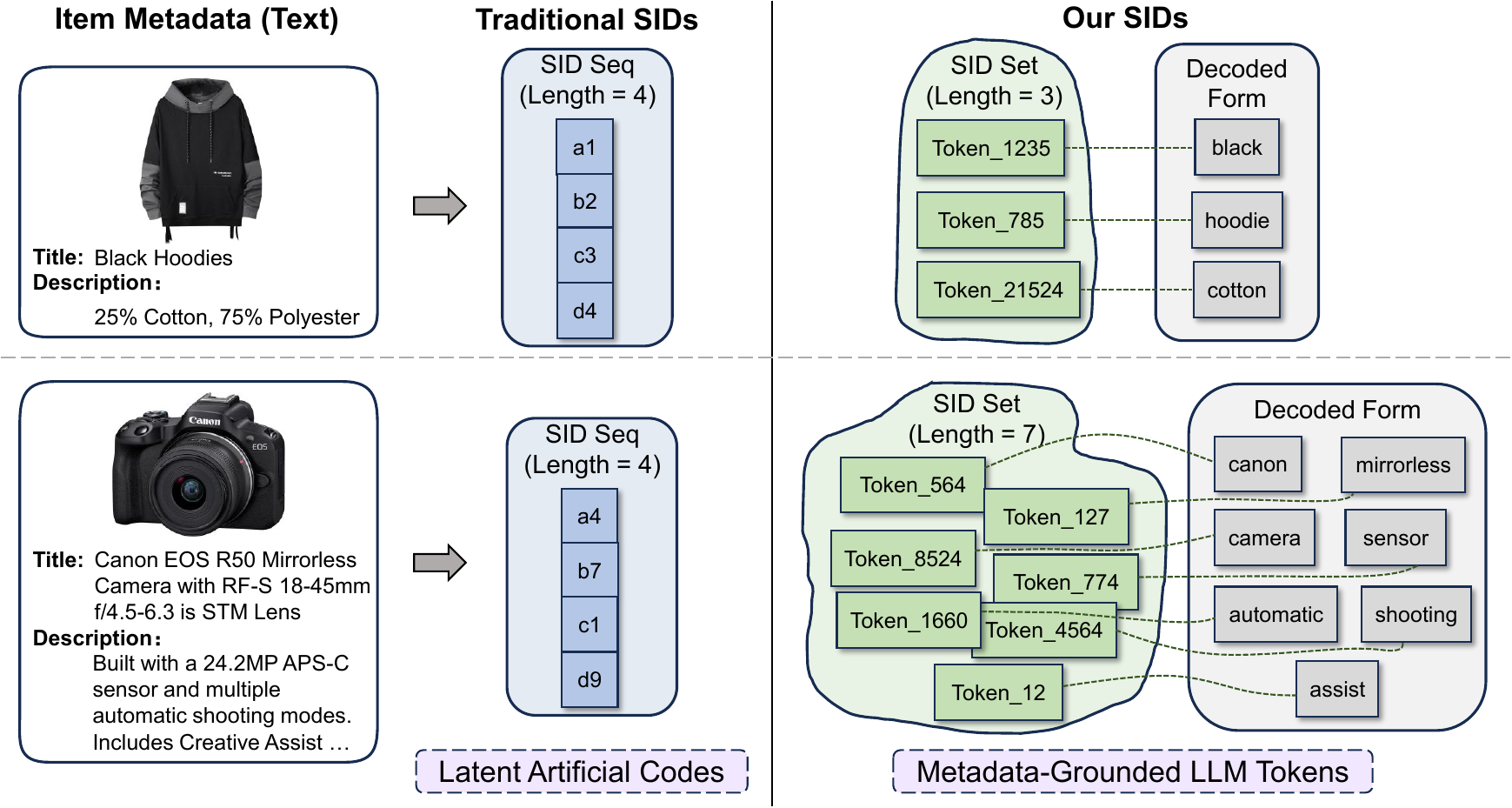}
    \caption{Comparison of traditional fixed-length SID sequences and our
    variable-length SID sets. Our SIDs are stored as LLM vocabulary token IDs, denoted as \texttt{Token\_ID} for illustration, while the decoded forms show their correspondence to item metadata.}
    \label{fig:sid_comparison}
    % \vspace{-10pt}
\end{figure}

However, existing SID-based methods still exhibit three major limitations. First, most SIDs are constructed by clustering or quantizing continuous item representations into artificial discrete codes~\cite{jia2025principles}. Although these codes may preserve useful latent information, they are neither explicitly grounded in the original item metadata nor naturally aligned with the pretrained LLM vocabulary. Consequently, their semantic content is difficult to inspect or quantify, and their integration often requires vocabulary expansion, newly initialized code embeddings, or additional task-specific alignment.
Second, most methods construct fixed-length SIDs~\cite{rajput2023recommender,hou2025generating}, assigning the same representational budget to all items regardless of their semantic content. Such a uniform design cannot adapt identifier capacity to item-specific needs: some items may be sufficiently represented by a few informative elements, whereas others may require richer identifiers to retain their distinguishing information.
Third, existing methods typically generate SIDs autoregressively with beam search, resulting in decoding costs that increase with both identifier length and beam size. This overhead becomes particularly pronounced when longer identifiers are needed to provide greater representational capacity.

To address these limitations, we propose VaLiDRec, a two-stage generative recommendation framework based on variable-length, LLM-aligned semantic identifiers. Rather than compressing item representations into artificial discrete codes, VaLiDRec constructs SIDs directly from the original item text using native LLM vocabulary tokens. In the first stage, it identifies semantically informative and discriminative tokens and employs a semantic-quality-aware greedy pruning strategy to select a compact token set that preserves the core meaning of each item. Consequently, each item is represented by a concise SID whose length naturally adapts to its semantic complexity. In the second stage, VaLiDRec incorporates collaborative signals through graph-aware behavioral prompts and fine-tunes the LLM with LoRA~\cite{hu2022lora} to predict the target item's SID. Training is jointly optimized with three complementary objectives: token-set prediction, semantic alignment, and recommendation ranking, which together promote accurate SID prediction, semantic consistency, and preference-aware inference. Finally, the predicted token scores are aggregated over all item SIDs
to produce item-level relevance scores, eliminating autoregressive SID generation and beam search.

In summary, our main contributions are as follows:
\begin{itemize}
    \item We propose a variable-length, LLM-aligned SID construction method that directly selects informative native LLM vocabulary tokens as item identifiers. Unlike quantization-based approaches, it avoids opaque latent codebooks, preserves explicit lexical semantics, and enables measurable semantic retention.

    \item We develop a graph-prompted token-set recommendation framework that injects collaborative signals into LLMs through graph-aware soft prompts. By predicting SID token sets in parallel and aggregating token scores over item SIDs, VaLiDRec naturally supports variable-length identifiers while eliminating autoregressive decoding and beam-search overhead.

    \item We conduct extensive experiments on multiple real-world recommendation datasets. The results demonstrate that VaLiDRec consistently outperforms both conventional and state-of-the-art sequential and SID-based generative recommendation baselines, while achieving superior inference efficiency, stronger zero-shot cold-start performance, and higher semantic retention.
\end{itemize}

\section{Related Work}
\subsection{Sequential Recommendation}

Sequential recommendation predicts the next item from interaction histories. FPMC~\cite{rendle2010factorizing} combines matrix factorization with first-order Markov chains for personalized transition modeling. Neural methods, including GRU4Rec~\cite{hidasi2015session}, NARM~\cite{li2017neural}, and Caser~\cite{tang2018personalized}, capture temporal dependencies, user intent, and local patterns, while SR-GNN~\cite{wu2019session} models sessions as item-transition graphs. Transformer-based methods further capture long-range dependencies, such as SASRec~\cite{kang2018self} with causal self-attention and BERT4Rec~\cite{sun2019bert4rec} with masked-item prediction. More recent work enhances sequence representations through self-supervision and contrastive learning, including S$^3$-Rec~\cite{zhou2020s3}, CL4SRec~\cite{xie2022contrastive}, and FEARec~\cite{du2023frequency}.

Despite their effectiveness, these methods primarily rely on item IDs and interaction signals within a discriminative ranking paradigm, limiting their use of item metadata and generalization to sparse or unseen items.

\subsection{Generative Recommendation}
GR retrieves items by generating identifiers from user histories. TIGER~\cite{rajput2023recommender} uses residual-quantized SIDs with autoregressive decoding, while LC-Rec~\cite{zheng2024adapting} aligns quantized SIDs with LLM representations. Later work improves identifier learning through collaborative tokenization, embedding alignment, or end-to-end optimization, including TokenRec~\cite{qu2025tokenrec}, LETTER~\cite{wang2024learnable}, ETEGRec~\cite{liu2025generative}, and DIGER~\cite{fu2026differentiable}. HSTU~\cite{zhai2024actions}, EAGER~\cite{wang2024eager}, and OneRec~\cite{deng2025onerec} further advance behavior modeling and unified retrieval-ranking generation.

Limited SID capacity has been identified as a bottleneck in generative recommendation~\cite{liu2025understanding}. RPG~\cite{hou2025generating} expands capacity with fixed-length OPQ codes, CapsID~\cite{cheng2026capsid} constructs variable-length latent SID sequences via soft capsule routing, and SA$^2$CRQ~\cite{wang2026towards} adaptively truncates RQ-VAE paths. Unlike these latent-code approaches, VaLiDRec constructs metadata-grounded SID sets directly from native LLM vocabulary tokens and predicts them in parallel.

\begin{figure*}[t]
    \centering
    \includegraphics[width=0.9\textwidth]{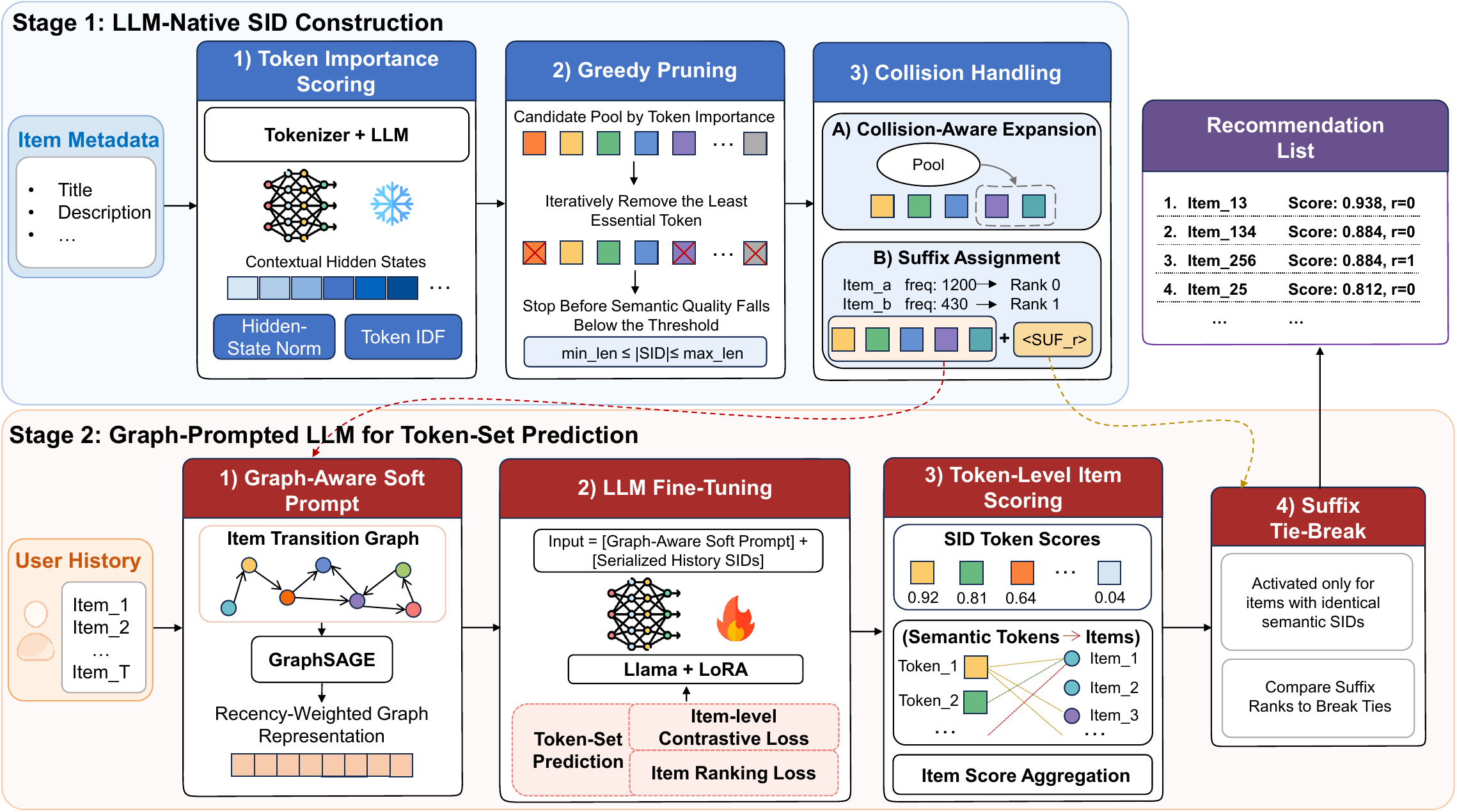}
    \caption{
    Overall framework of VaLiDRec. 
    }
    \label{fig:framework}
    % \vspace{-10pt}
\end{figure*}

\section{Problem Formulation}

Let $\mathcal{U}$ and $\mathcal{I}$ denote the user and item sets.
For each user $u\in\mathcal{U}$, let
$\mathcal{H}_u=(i_{u,1},\ldots,i_{u,T_u})$
denote the chronologically ordered interaction history.
Given $\mathcal{H}_u$, the sequential recommendation task is to predict the next item $j_u^+\in\mathcal{I}$.

Each item $j\in\mathcal{I}$ is associated with metadata $x_j$.
From $x_j$, we construct a variable-length semantic token set
$S_j\subseteq\mathcal{V}$, where $\mathcal{V}$ denotes the pretrained
LLM vocabulary. We formulate next-item recommendation as predicting
the semantic token set of the target item:
\begin{equation}
p_{\theta}\left(S_{j_u^+}\mid\mathcal{H}_u\right).
\end{equation}
At inference time, the predicted token scores are aggregated over the SIDs of all candidate items to obtain item-level relevance scores and rank the full item catalog.

\section{Methodology}
In this section, we present VaLiDRec, a two-stage GR framework based on variable-length, LLM-aligned SIDs. As illustrated in
Figure~\ref{fig:framework}, Stage~1 (Section~\ref{sec:stage1}) constructs semantic SIDs directly from item metadata, while Stage~2 (Section~\ref{sec:stage2}) incorporates collaborative signals through graph-aware soft prompts and performs parallel token-set prediction for item ranking.

\subsection{Stage 1: LLM-Native SID Construction}
\label{sec:stage1}

Existing SID methods use clustering or quantization with fixed code structures, assigning equal capacity to all items and potentially losing fine-grained semantics. Their latent codes are also misaligned with item metadata and pretrained LLM vocabularies. We instead construct variable-length SIDs from native LLM tokens, adapting their capacity to each item's semantic complexity.

\subsubsection{Token Importance Scoring}

Given the metadata text $x_j$ of item $j$, the LLM tokenizer produces
\begin{equation}
T_j=(t_{j,1},\ldots,t_{j,n_j}),
\end{equation}
where $n_j$ is the sequence length. Let
$\mathcal{P}_j\subseteq\{1,\ldots,n_j\}$ denote the positions of valid
metadata tokens after filtering, and let
$\mathbf{h}_{j,i}\in\mathbb{R}^{d}$ be the final-layer contextual
representation at position $i\in\mathcal{P}_j$.

To identify tokens that are both salient to the current item and
discriminative across the corpus, we combine contextual activation
with inverse document frequency:
\begin{equation}
\operatorname{IDF}(t)
=
\log
\frac{|\mathcal{I}|+1}
{\operatorname{df}(t)+1},
\end{equation}
where $\operatorname{df}(t)$ denotes the number of items whose metadata
contains token type $t$.

Unlike conventional TF-IDF~\cite{salton1988term}, which uses term
frequency as the document-specific importance signal, we use the
contextual hidden-state norm as a proxy for token salience. The
occurrence-level importance score is
\begin{equation}
s_{j,i}
=
\|\mathbf{h}_{j,i}\|_2
\left(
\operatorname{IDF}(t_{j,i})+1
\right),
\qquad i\in\mathcal{P}_j.
\end{equation}
This score favors token occurrences that are both contextually salient
and corpus-discriminative. We construct a candidate position set $\mathcal{C}_j \subseteq \mathcal{P}_j$ by combining high-ranking candidates under
$s_{j,i}$, $\mathrm{IDF}(t_{j,i})$, and $\|\mathbf{h}_{j,i}\|_2$, together with a small number of prefix positions. If multiple positions correspond to the same vocabulary token, we retain only the occurrence with the highest importance score.

\subsubsection{Greedy Pruning}

The candidate pool may contain redundant token occurrences, whereas retaining a fixed number of tokens assigns the same representational capacity to items with different semantic complexity. We therefore introduce a semantic-quality-aware greedy pruning strategy to construct compact, variable-length SIDs.

We represent the metadata semantics of item $j$ as
\begin{equation}
\mathbf{e}_j
=
\frac{1}{|\mathcal{P}_j|}
\sum_{i\in\mathcal{P}_j}
\mathbf{h}_{j,i}.
\end{equation}
Let $\mathcal{R}_j\subseteq\mathcal{C}_j$ denote the currently retained
token positions. Their semantic representation is
\begin{equation}
\mathbf{e}(\mathcal{R}_j)
=
\frac{1}{|\mathcal{R}_j|}
\sum_{i\in\mathcal{R}_j}
\mathbf{h}_{j,i},
\end{equation}
and the corresponding semantic quality is
\begin{equation}
Q(\mathcal{R}_j)
=
\cos\left(
\mathbf{e}(\mathcal{R}_j),
\mathbf{e}_j
\right).
\end{equation}
A larger $Q(\mathcal{R}_j)$ indicates better preservation of the metadata semantics.

We initialize $\mathcal{R}_j\leftarrow\mathcal{C}_j$ and iteratively
remove the least essential token occurrence. Given a trade-off
coefficient $\lambda\geq0$, the deletion score of position
$i\in\mathcal{R}_j$ is
\begin{equation}
D_j(i;\mathcal{R}_j)
=
Q\left(\mathcal{R}_j\setminus\{i\}\right)
-
\lambda
\frac{
[\operatorname{IDF}(t_{j,i})]_+
}{
\sum_{i'\in\mathcal{R}_j}
[\operatorname{IDF}(t_{j,i'})]_+
+\epsilon
},
\end{equation}
where $[x]_+=\max(x,0)$ and $\epsilon>0$ is a small constant for
numerical stability. The first term favors deletions that preserve
item semantics, whereas the second penalizes the removal of
corpus-discriminative tokens. At each iteration, we select
\begin{equation}
i^\star
=
\operatorname*{arg\,max}_{i\in\mathcal{R}_j}
D_j(i;\mathcal{R}_j).
\end{equation}

The pruning process is controlled by a minimum length $L_{\min}$, a
maximum length $L_{\max}$, and a semantic-quality threshold $\tau$.
If $|\mathcal{R}_j|>L_{\max}$, $i^\star$ is removed to satisfy the hard
maximum-length constraint. Once $|\mathcal{R}_j|\leq L_{\max}$,
removal continues only if $|\mathcal{R}_j|>L_{\min}$ and
\begin{equation}
Q\left(\mathcal{R}_j\setminus\{i^\star\}\right)
\geq\tau.
\end{equation}
The process terminates otherwise. Thus, semantically simple items
receive shorter identifiers, whereas information-rich items retain
more tokens. 
Since the candidate positions correspond to distinct token IDs, the resulting semantic SID is
\begin{equation}
\mathcal{S}_j
=
\left\{
t_{j,i} \mid i \in \mathcal{R}_j
\right\}.
\end{equation}
After each update to $\mathcal{R}_j$, the semantic SID $S_j$ is updated accordingly.

\subsubsection{Collision Handling}

Greedy pruning may assign identical semantic token sets to items with
similar metadata. We therefore perform collision-aware refinement while
preserving token occurrences that contribute most to semantic
retention.

For a retained position $i\in\mathcal{R}_j$, we define its functional
contribution as
\begin{equation}
\Delta_j(i)
=
\left[
Q(\mathcal{R}_j)
-
Q\left(\mathcal{R}_j\setminus\{i\}\right)
\right]_+.
\end{equation}
A larger $\Delta_j(i)$ indicates a greater loss of semantic quality
after removing the corresponding token. Its normalized functional
weight is
\begin{equation}
w_j(i)
=
\begin{cases}
\dfrac{\Delta_j(i)}
{\sum_{i'\in\mathcal{R}_j}\Delta_j(i')},
&
\text{if }
\sum_{i'\in\mathcal{R}_j}\Delta_j(i')>0,
\\[8pt]
\dfrac{s_{j,i}}
{\sum_{i'\in\mathcal{R}_j}s_{j,i'}},
&
\text{otherwise}.
\end{cases}
\end{equation}
The second branch provides a fallback when removing any individual
token does not reduce semantic quality.

Two items $j$ and $j'$ are considered collided if they share the same
semantic SID:
\begin{equation}
S_j=S_{j'}.
\end{equation}
For each collided item, we select an unused candidate position
$i_{\mathrm{new}} \in \mathcal{C}_j \setminus \mathcal{R}_j$
whose token $t_{j,i_{\mathrm{new}}}$ is not already contained in
$\mathcal{S}_j$, prioritizing larger $\mathrm{IDF}(t_{j,i_{\mathrm{new}}})$
and using $s_{j,i_{\mathrm{new}}}$ to break ties.
If $|\mathcal{R}_j| < L_{\max}$, the selected position is added to
$\mathcal{R}_j$; otherwise, it replaces the retained position with the
smallest functional weight. Collision groups are recomputed after each
round, and refinement continues until all collisions are resolved or
the maximum number of rounds is reached.

Although $S_j$ is unordered, we deterministically serialize it as
$\pi(S_j)$ by sorting tokens by decreasing stored weight and
occurrence-level importance. For each residual collision group, items are ranked by training interaction frequency, with ties broken by a fixed item index, yielding ranks~$r_j$ from zero. Using $\oplus$ to denote concatenation, the complete identifier is
\begin{equation}
\widetilde{S}_j
=
\pi(S_j)
\oplus
\langle\mathrm{SUF}_{r_j}\rangle.
\label{eq:suffix_assignment}
\end{equation}
Non-collided items use $\langle\mathrm{SUF}_0\rangle$. As a non-semantic disambiguation key, the suffix is excluded from graph construction, LLM input, token prediction, and semantic scoring. Hereafter, $S_j$ and $\widetilde{S}_j$ denote the semantic token set and complete stored identifier, respectively.

\subsection{Stage 2: Graph-Prompted LLM for Token-Set Prediction}
\label{sec:stage2}

The semantic SIDs from Section~\ref{sec:stage1} capture item content but not behavioral transitions. Moreover, autoregressive decoding introduces artificial token ordering and costly beam search for variable-length token sets. We therefore formulate recommendation as graph-conditioned token-set prediction, where a graph-aware soft prompt models transition patterns and the LLM predicts SID tokens in parallel. Token scores are then aggregated across item SIDs to obtain item-level recommendation scores.

\subsubsection{Graph-Aware Soft Prompt}

Let $\mathbf{E}(t)\in\mathbb{R}^{d}$ denote the LLM input embedding of
token $t$. We initialize the representation of item $j$ by averaging
the embeddings of its semantic SID tokens:
\begin{equation}
\mathbf{z}_j
=
\frac{1}{|S_j|}
\sum_{t\in S_j}
\mathbf{E}(t).
\end{equation}

For each mini-batch $\mathcal{B}$, we construct a directed
item-transition graph from the user histories. For each user
$u$, we add an edge
$i_{u,\ell}\rightarrow i_{u,\ell+1}$ between consecutive interactions
in $\mathcal{H}_u$, together with self-loops for historical items.

We initialize $\mathbf{g}_j^{(0)}=\mathbf{z}_j$. Let
$\mathcal{N}^{-}(j)$ denote the multiset of incoming neighbors of item
$j$. Each GraphSAGE~\cite{hamilton2017inductive} layer updates the node representation as
\begin{equation}
\mathbf{g}_j^{(l+1)}
=
\mathrm{ReLU}\left(
\mathbf{W}^{(l)}
\left[
\mathbf{g}_j^{(l)}
\Vert
\frac{1}{|\mathcal{N}^{-}(j)|}
\sum_{v\in\mathcal{N}^{-}(j)}
\mathbf{g}_v^{(l)}
\right]
+
\mathbf{b}^{(l)}
\right),
\end{equation}
where $l=0,\ldots,L-1$,
$\mathbf{W}^{(l)}\in\mathbb{R}^{d\times 2d}$,
$\mathbf{b}^{(l)}\in\mathbb{R}^{d}$, and $\Vert$ denotes
concatenation.

We summarize the graph-enhanced historical item representations using
recency-weighted pooling:
\begin{equation}
\mathbf{r}_u
=
\sum_{\ell=1}^{T_u}
\rho_{u,\ell}
\mathbf{g}^{(L)}_{i_{u,\ell}},
\qquad
\rho_{u,\ell}
=
\frac{\ell}{\sum_{q=1}^{T_u}q}.
\end{equation}

The pooled representation is projected into $P$ continuous prompt
embeddings:
\begin{equation}
\mathbf{P}_u
=
\mathrm{reshape}
\left(
\mathbf{W}_p\mathbf{r}_u+\mathbf{b}_p
\right)
\in\mathbb{R}^{P\times d},
\end{equation}
where $\mathbf{W}_p\in\mathbb{R}^{Pd\times d}$ and
$\mathbf{b}_p\in\mathbb{R}^{Pd}$. Each row of $\mathbf{P}_u$ is a $d$-dimensional prompt embedding compatible with the LLM input space.

Finally, we serialize and concatenate the historical semantic SIDs as
$\pi(S_{i_{u,1}})\oplus\cdots\oplus\pi(S_{i_{u,T_u}})$ and prepend
$\mathbf{P}_u$ to their token embeddings. Auxiliary suffix tokens are
excluded from both graph construction and the LLM input.

\subsubsection{LLM Fine-Tuning}

We fine-tune the LLM using LoRA together with the graph-prompt module. Let $\mathcal{V}_{\mathrm{SID}}$ denote the set of all semantic tokens appearing in the item SIDs, excluding auxiliary suffix tokens. During training, gradients of the input embedding matrix are restricted to tokens in $\mathcal{V}_{\mathrm{SID}}$, while all other token embeddings remain frozen. The training objective combines token-set prediction, item-level ranking, and contrastive alignment.

Let $\mathbf{h}_u\in\mathbb{R}^{d}$ be the final-layer hidden state at the designated prediction position following the serialized history after prepending the graph-aware soft prompt.
To score SID tokens in the LLM vocabulary space, we reuse the
corresponding input embeddings as output classifiers. For each
$t\in\mathcal{V}_{\mathrm{SID}}$, we compute
\begin{equation}
o_{u,t}
=
\mathbf{h}_u^{\top}\mathbf{E}(t),
\end{equation}
where $\mathbf{E}(t)$ is the corresponding LLM input embedding.

Given a mini-batch $\mathcal{B}$, we define the candidate-token set as
\begin{equation}
\mathcal{T}_{\mathcal{B}}
=
\bigcup_{u\in\mathcal{B}} S_{j_u^+}.
\end{equation}
For user $u$, tokens in $S_{j_u^+}$ are treated as positives and the remaining tokens in $\mathcal{T}_{\mathcal{B}}$ as in-batch negatives.
Let $y_{u,t}=\mathbb{I}[t\in S_{j_u^+}]$. The token-set prediction loss
is
\begin{equation}
\begin{aligned}
\mathcal{L}_{\mathrm{token}}
=
&-\frac{1}
{|\mathcal{B}|\,|\mathcal{T}_{\mathcal{B}}|}
\sum_{u\in\mathcal{B}}
\sum_{t\in\mathcal{T}_{\mathcal{B}}}
\Big[
\omega y_{u,t}\log\sigma(o_{u,t})
\\
&\qquad+
(1-y_{u,t})
\log\bigl(1-\sigma(o_{u,t})\bigr)
\Big],
\end{aligned}
\end{equation}
where $\omega>0$ is the positive-class weight.

Token-level supervision captures target SID tokens but does not directly optimize item-level ranking. We therefore introduce an item-ranking objective that promotes the positive item over sampled negatives. 
For each user, we construct
$\mathcal{A}_u=\{j_u^+\}\cup\mathcal{A}_u^-$, where
$\mathcal{A}_u^-$ contains uniformly sampled negative items. 

The score of candidate item $j$ is
\begin{equation}
a_{u,j}
=
\frac{1}{|S_j|}
\sum_{t\in S_j}
\sigma(o_{u,t}),
\end{equation}
and the item-ranking loss is
\begin{equation}
\mathcal{L}_{\mathrm{rank}}
=
-\frac{1}{|\mathcal{B}|}
\sum_{u\in\mathcal{B}}
\log
\frac{
\exp(a_{u,j_u^+})
}{
\sum_{j\in\mathcal{A}_u}\exp(a_{u,j})
}.
\end{equation}
Although the token-set and ranking objectives optimize token prediction and item ordering, they do not explicitly align the LLM-derived user representation with graph-enhanced item representations. We therefore introduce an item-level contrastive objective that aligns each user representation with its target-item representation relative to other target representations in the mini-batch.
For contrastive alignment, let
\begin{equation}
\bar{\mathbf{g}}_j
=
\begin{cases}
\mathbf{g}_j^{(L)},
&
\text{if $j$ appears in the mini-batch graph},
\\
\mathbf{z}_j,
&
\text{otherwise}.
\end{cases}
\end{equation}
We align the user representation with the graph-enhanced representation of its target item using
\begin{equation}
\begin{aligned}
\mathcal{L}_{\mathrm{item}}
=
-\frac{1}{|\mathcal{B}|}
\sum_{u\in\mathcal{B}}
\log
\frac{
\exp\left(
\mathrm{sim}(\mathbf{h}_u,\bar{\mathbf{g}}_{j_u^+})/\tau_c
\right)
}{
\sum_{v\in\mathcal{B}}
\exp\left(
\mathrm{sim}(\mathbf{h}_u,\bar{\mathbf{g}}_{j_v^+})/\tau_c
\right)
},
\end{aligned}
\end{equation}
where $\mathrm{sim}(\cdot,\cdot)$ denotes cosine similarity and
$\tau_c$ is the contrastive temperature. This objective aligns the LLM-derived user representation with the content- and graph-enhanced target-item space.

Finally, the overall training objective is
\begin{equation}
\mathcal{L}
=
\mathcal{L}_{\mathrm{token}}
+
\alpha\mathcal{L}_{\mathrm{rank}}
+
\beta\mathcal{L}_{\mathrm{item}},
\end{equation}
where $\alpha$ and $\beta$ control the item-ranking and contrastive
objectives, respectively.

\subsubsection{Token-Level Item Scoring}

The fine-tuned LLM computes all semantic SID token scores in parallel,
rather than autoregressively generating complete identifiers. For each
token $t\in\mathcal{V}_{\mathrm{SID}}$, we compute
\begin{equation}
p_{u,t}
=
\sigma(o_{u,t})
=
\sigma\left(
\mathbf{h}_u^{\top}\mathbf{E}(t)
\right).
\end{equation}

The relevance score of item $j$ is obtained by averaging the
probabilities of its semantic SID tokens:
\begin{equation}
q_{u,j}
=
\frac{1}{|S_j|}
\sum_{t\in S_j}
p_{u,t}.
\end{equation}
The final recommendation list is obtained by ranking all items
$j\in\mathcal{I}$ according to $q_{u,j}$. The length normalization
prevents items with longer SIDs from systematically accumulating larger scores. Since all token probabilities are computed in a single LLM forward pass, this scoring process eliminates autoregressive SID generation and beam-search decoding.

\subsubsection{Suffix Tie-Break}

Collision-aware expansion resolves most SID collisions, but some items
may still share identical semantic token sets. Since suffix labels carry no semantic information, they are excluded from graph construction, LLM input, token prediction, and semantic score aggregation, and are used only for deterministic disambiguation.

Consider two candidate items $j_a,j_b\in\mathcal{I}$ such that
$S_{j_a}=S_{j_b}$. Their semantic scores are identical under
token-level aggregation, so the item with the smaller suffix rank is
preferred:
\begin{equation}
S_{j_a}=S_{j_b}
\quad\Longrightarrow\quad
j_a\succ j_b
\iff
r_{j_a}<r_{j_b}.
\end{equation}
Thus, suffix ranks affect only items with identical semantic token sets and do not alter any strict semantic-score ordering.

\begin{table}[t]
\centering
\caption{Statistics of the standard and zero-shot item cold-start datasets.}
\label{tab:dataset_statistics}
\resizebox{0.9\linewidth}{!}{
\begin{tabular}{l|cccc}
\hline
\textbf{Dataset} 
& \textbf{\#Users} 
& \textbf{\#Items} 
& \textbf{\#Interactions} 
& \textbf{Sparsity} \\
\hline
\multicolumn{5}{c}{\textbf{Standard Evaluation (5-core)}} \\
\hline
\textit{Luxury}      & 1,841  & 842   & 18,667  & 98.80\% \\
\textit{Scientific}  & 2,843  & 1,549 & 19,099  & 99.57\% \\
\textit{Instruments} & 17,112 & 6,250 & 136,226 & 99.87\% \\
\textit{Arts}        & 22,171 & 9,416 & 174,079 & 99.92\% \\
\hline
\multicolumn{5}{c}{\textbf{Zero-Shot Item Cold-Start Evaluation}} \\
\hline
\textit{Luxury} & 19,748 & 8,308 & 47,913 & 99.97\% \\
\hline
\end{tabular}
}
\end{table}

\section{Experiments}
\subsection{Datasets and Evaluation Metrics}

We conduct standard experiments on four Amazon-2018 domains: Luxury Beauty, Industrial and Scientific, Musical Instruments, and Arts, Crafts and Sewing. We retain items with available metadata and apply iterative 5-core filtering, ensuring that each remaining user and item has at least five interactions. For item cold-start evaluation, following Zhang et al.~\cite{zhang2026cold}, we construct an item-cold split from the raw Luxury
Beauty data without 5-core filtering. For each user sequence, the first 90\% of interactions are used to construct training instances, while unseen items from the remaining interactions are selected as cold targets. The resulting cold-start instances are evenly divided into validation and test sets.

We report Recall@K and Normalized Discounted Cumulative Gain (NDCG@K), which measure top-$K$ target coverage and ranking quality, respectively. For standard evaluation, we use $K\in\{5,10,20\}$; for zero-shot item cold-start evaluation, we report results at $K\in\{50,100\}$ due to the substantially larger and sparser candidate space.

\subsection{Implementation Details and Hyperparameter Settings}
All experiments are conducted on NVIDIA H100 GPUs. We use Llama-3.1-8B for SID construction and Llama-3.2-1B for recommendation training~\cite{grattafiori2024llama}; the two models can be replaced by any pair with compatible tokenizers and vocabulary mappings. For SID construction, we set the maximum metadata length, semantic quality threshold, and candidate pool size to 2,048, 0.95, and 64, respectively. SIDs contain 2--16 semantic tokens, with a distinctiveness coefficient of 0.05 and up to eight collision-aware expansion rounds.  For recommendation, we use a two-layer GraphSAGE encoder with eight soft prompt embeddings and the most recent 20 historical items. We apply LoRA to the query and value projections with rank 16, scaling factor 32, and dropout 0.05. We optimize using AdamW with a learning rate of $1\times10^{-4}$ and a batch size of 32. We set $\beta=0.1$ and analyze $\alpha$ in Section~\ref{sec:hyperparameter}. 
For efficiency evaluation, both VaLiDRec and LC-Rec use Llama-3.2-1B and are tested on the same NVIDIA H100 GPU with a batch size of 1; further details are provided in Section~\ref{sec:efficiency}.
All results are averaged over five independent runs with different
random seeds. Statistical significance is assessed using a paired
$t$-test at $p<0.05$.

\subsection{Baselines}

We compare VaLiDRec with two groups of representative baselines, including traditional SR methods and recent GR methods.

\begin{itemize}
    \item \textbf{GRU4Rec}~\cite{hidasi2015session} models sequential preferences with gated recurrent units.

    \item \textbf{Caser}~\cite{tang2018personalized} captures sequential patterns using convolution over recent interactions.

    \item \textbf{SASRec}~\cite{kang2018self} models long-range dependencies with self-attention.

    \item \textbf{BERT4Rec}~\cite{sun2019bert4rec} adopts a bidirectional Transformer architecture and learns user sequential preferences through a masked item prediction objective.

    \item \textbf{TIGER}~\cite{rajput2023recommender} is a GR method that constructs residual-quantized SIDs and uses T5~\cite{raffel2020exploring} to autoregressively generate target item identifiers.

    \item \textbf{LC-Rec}~\cite{zheng2024adapting} is an LLM-based GR method that leverages language models to enhance item representation and recommendation generation.

    \item \textbf{RPG}~\cite{hou2025generating} is a recent GR method that improves recommendation by incorporating semantic and preference-aware generation signals.

    \item \textbf{SA$^2$CRQ}~\cite{wang2026towards} is a GR method that improves long-tail recommendation by learning adaptive semantic codes and transferring residual knowledge across items.\footnote{As the original implementation has not been publicly released, we reproduce SA$^2$CRQ based on the method described in the paper.}
\end{itemize}

\subsection{Performance Study}
\begin{table*}[t]
\centering
\caption{Overall recommendation performance on four datasets. The best results are highlighted in bold, and the second-best results are underlined.}
\label{tab:overall_results}
\resizebox{0.85\linewidth}{!}{
\begin{tabular}{llccccccccc}
\toprule
\textbf{Dataset} & \textbf{Metric}
& \textbf{GRU4Rec}
& \textbf{Caser}
& \textbf{SASRec}
& \textbf{BERT4Rec}
& \textbf{TIGER}
& \textbf{LC-Rec}
& \textbf{RPG}
& \textbf{SA$^2$CRQ}
& \textbf{VaLiDRec} \\
\midrule

\multirow{6}{*}{\textit{Luxury}}
& Recall@5  & 0.2846 & 0.1879 & 0.2042 & 0.2330 & 0.2629 & 0.2622 & \underline{0.2906} & 0.2542 & \textbf{0.3014} \\
& Recall@10 & 0.3128 & 0.2927 & 0.3014 & 0.3096 & 0.2826 & 0.2772 & \underline{0.3167} & 0.2716 & \textbf{0.3405} \\
& Recall@20 & 0.3530 & 0.3324 & 0.3476 & 0.3531 & 0.3147 & 0.2955 & \underline{0.3552} & 0.2944 & \textbf{0.3992} \\
& NDCG@5    & 0.2355 & 0.1127 & 0.1703 & 0.1465 & 0.2457 & 0.2421 & \underline{0.2512} & 0.2408 & \textbf{0.2597} \\
& NDCG@10   & 0.2434 & 0.1465 & 0.1879 & 0.1719 & 0.2518 & 0.2470 & \underline{0.2590} & 0.2462 & \textbf{0.2709} \\
& NDCG@20   & 0.2547 & 0.1565 & 0.1909 & 0.1828 & 0.2608 & 0.2516 & \underline{0.2691} & 0.2519 & \textbf{0.2859} \\
\midrule

\multirow{6}{*}{\textit{Scientific}}
& Recall@5  & \underline{0.0661} & 0.0422 & 0.0292 & 0.0496 & 0.0555 & 0.0300 & 0.0531 & 0.0337 & \textbf{0.0685} \\
& Recall@10 & \underline{0.0893} & 0.0647 & 0.0528 & 0.0735 & 0.0755 & 0.0489 & 0.0711 & 0.0502 & \textbf{0.0942} \\
& Recall@20 & \underline{0.1259} & 0.1093 & 0.0837 & 0.1136 & 0.1095 & 0.0649 & 0.1006 & 0.0731 & \textbf{0.1287} \\
& NDCG@5    & \underline{0.0466} & 0.0256 & 0.0166 & 0.0329 & 0.0370 & 0.0181 & 0.0341 & 0.0228 & \textbf{0.0475} \\
& NDCG@10   & \underline{0.0541} & 0.0325 & 0.0240 & 0.0405 & 0.0434 & 0.0241 & 0.0386 & 0.0285 & \textbf{0.0552} \\
& NDCG@20   & \underline{0.0617} & 0.0432 & 0.0317 & 0.0506 & 0.0520 & 0.0283 & 0.0465 & 0.0340 & \textbf{0.0637} \\
\midrule

\multirow{6}{*}{\textit{Instruments}}
& Recall@5  & 0.0797 & 0.0570 & 0.0427 & 0.0694 & \underline{0.0839} & 0.0559 & 0.0759 & 0.0771 & \textbf{0.0922} \\
& Recall@10 & 0.0906 & 0.0797 & 0.0637 & 0.0871 & \underline{0.0946} & 0.0659 & 0.0806 & 0.0888 & \textbf{0.1042} \\
& Recall@20 & 0.1087 & 0.1074 & 0.0924 & \underline{0.1107} & 0.1087 & 0.0874 & 0.0877 & 0.1055 & \textbf{0.1217} \\
& NDCG@5    & 0.0712 & 0.0348 & 0.0240 & 0.0603 & \underline{0.0773} & 0.0446 & 0.0678 & 0.0687 & \textbf{0.0837} \\
& NDCG@10   & 0.0747 & 0.0421 & 0.0304 & 0.0660 & \underline{0.0807} & 0.0478 & 0.0694 & 0.0726 & \textbf{0.0875} \\
& NDCG@20   & 0.0792 & 0.0490 & 0.0378 & 0.0721 & \underline{0.0842} & 0.0533 & 0.0711 & 0.0768 & \textbf{0.0917} \\
\midrule

\multirow{6}{*}{\textit{Arts}}
& Recall@5  & \underline{0.0658} & 0.0330 & 0.0607 & 0.0576 & 0.0587 & 0.0194 & 0.0538 & 0.0476 & \textbf{0.0803} \\
& Recall@10 & 0.0770 & 0.0472 & \underline{0.0852} & 0.0782 & 0.0678 & 0.0258 & 0.0583 & 0.0565 & \textbf{0.0934} \\
& Recall@20 & 0.0912 & 0.0672 & \underline{0.1059} & 0.1044 & 0.0806 & 0.0321 & 0.0645 & 0.0650 & \textbf{0.1110} \\
& NDCG@5    & \underline{0.0564} & 0.0199 & 0.0320 & 0.0398 & 0.0513 & 0.0143 & 0.0453 & 0.0408 & \textbf{0.0699} \\
& NDCG@10   & \underline{0.0600} & 0.0244 & 0.0398 & 0.0464 & 0.0541 & 0.0164 & 0.0468 & 0.0437 & \textbf{0.0739} \\
& NDCG@20   & \underline{0.0636} & 0.0294 & 0.0475 & 0.0530 & 0.0570 & 0.0179 & 0.0483 & 0.0458 & \textbf{0.0779} \\

\bottomrule
\end{tabular}
}
% \vspace{-10pt}
\end{table*}

We compare VaLiDRec with eight representative baselines, including four sequential recommenders and four generative SID-based methods.
Table~\ref{tab:overall_results} shows that VaLiDRec achieves the best
performance across all datasets, metrics, and cutoff values.
Conventional sequential models remain competitive because 5-core
filtering retains relatively active users and items, allowing reliable
behavioral patterns to be learned.

On the relatively dense Luxury dataset, VaLiDRec improves Recall@10 and Recall@20 over RPG, the strongest baseline, by \(7.5\%\) and \(12.4\%\), respectively. These results suggest that combining variable-length LLM-native SIDs with graph-enhanced preference modeling is more effective than the globally fixed code design used by RPG. On Scientific, VaLiDRec improves Recall@10 and NDCG@10 over GRU4Rec by \(5.5\%\) and \(2.0\%\), respectively, demonstrating its benefit even when sequential signals are strong. The advantage becomes more pronounced on the sparser datasets: VaLiDRec improves all Recall metrics on Instruments by approximately \(10\%\), and all NDCG metrics on Arts by more than \(22\%\). These gains indicate improved candidate coverage as well as more accurate ranking of relevant items at top positions.

TIGER and LC-Rec rely on quantized SIDs, whereas RPG uses globally fixed OPQ codes and SA$^2$CRQ adaptively truncates RQ-VAE paths. Although these methods differ in their quantization strategies, they all operate on artificial code spaces. By contrast, VaLiDRec constructs variable-length SIDs directly from the native vocabulary of a pretrained LLM. Collectively, the results highlight the importance of jointly modeling SID semantics, vocabulary alignment, and user behavioral transitions.

\subsubsection{Cold-Start Study}

\begin{table}[t]
\centering
\caption{Zero-shot item cold-start performance on Luxury. The best and
second-best results are highlighted in bold and underlined,
respectively.}
\label{tab:cold_start_results}
\resizebox{1.0\linewidth}{!}{
\begin{tabular}{lcccc}
\toprule
\textbf{Method}
& \textbf{Recall@50}
& \textbf{NDCG@50}
& \textbf{Recall@100}
& \textbf{NDCG@100} \\
\midrule
GRU4Rec
& 0.0000 & 0.0000
& 0.0000 & 0.0000 \\

BERT4Rec
& 0.0000 & 0.0000
& 0.0000 & 0.0000 \\

TIGER
& 0.0042 & 0.0008
& 0.0065 & 0.0012 \\

LC-Rec
& 0.1335 & \underline{0.0562}
& 0.1553 & 0.0619 \\

RPG
& \underline{0.1899} & 0.0555
& \underline{0.2509} & \underline{0.0640} \\

VaLiDRec
& \textbf{0.1975} & \textbf{0.0575}
& \textbf{0.2571} & \textbf{0.0672} \\
\bottomrule
\end{tabular}
}
% \vspace{-10pt}
\end{table}

Table~\ref{tab:cold_start_results} reports the zero-shot item
cold-start results on Luxury, where target items are absent from the
training interactions. GRU4Rec and BERT4Rec obtain zero performance because they cannot learn interaction-based representations for unseen target items.

Generative SID methods leverage item content but differ in identifier construction. TIGER uses RQ-VAE to construct residual-quantized SIDs and trains T5 for autoregressive generation, whereas LC-Rec aligns quantized SID tokens with an extended LLM vocabulary through additional fine-tuning. RPG instead adopts OPQ to construct globally fixed long SIDs from independently quantized subvectors and predicts them in parallel, yielding strong recall. Nevertheless, all these methods rely on artificial quantization codes. VaLiDRec improves Recall@50, NDCG@50, Recall@100, and NDCG@100 over
the corresponding strongest baselines by $4.0\%$, $2.3\%$, $2.5\%$,
and $5.0\%$, respectively. By constructing variable-length SIDs directly from native LLM vocabulary tokens, VaLiDRec represents unseen items from metadata in the pretrained semantic space. The gains, particularly in NDCG, demonstrate its ability to rank relevant cold items more accurately.

\subsection{Ablation Study}
\begin{figure*}[t]
    \centering
    \includegraphics[width=1.0\textwidth]{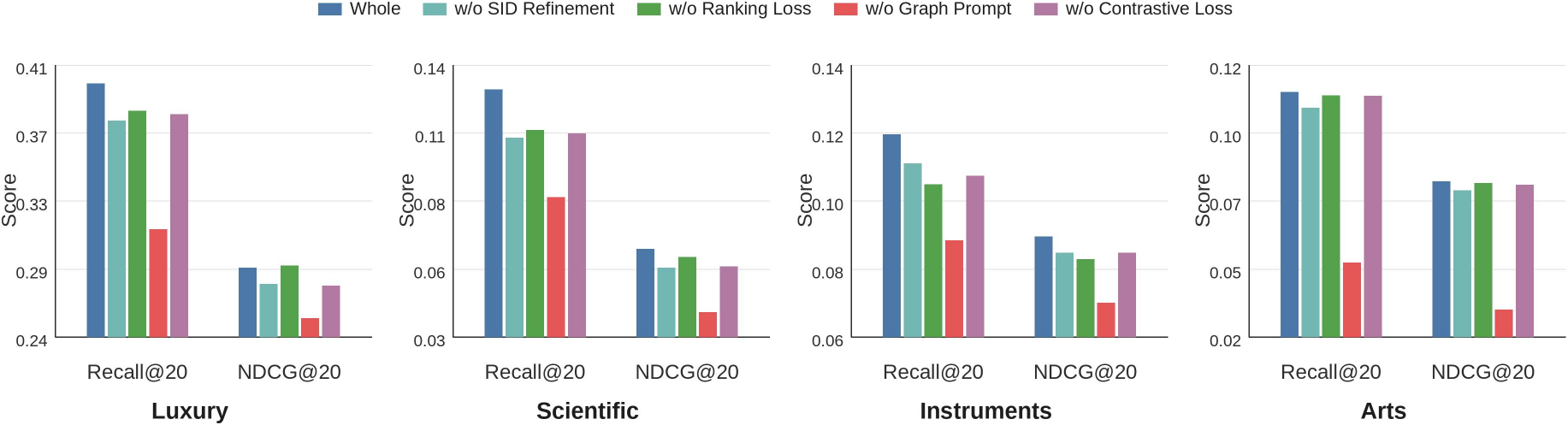}
    \caption{Ablation results on four datasets.}
    \label{fig:ablation}
    % \vspace{-10pt}
\end{figure*}

To assess each component, we consider four variants:
\textbf{w/o SID Refinement}, which removes greedy pruning and
collision-aware expansion;
\textbf{w/o Graph Prompt}, which removes the graph-aware soft prompt;
\textbf{w/o Ranking Loss}, which removes the item-level ranking loss;
and \textbf{w/o Contrastive Loss}, which removes the item-level
InfoNCE objective.

Figure~\ref{fig:ablation} reports Recall@20 and NDCG@20 on four
datasets. Removing the graph-aware soft prompt causes the largest
performance drop, especially on Arts, confirming its central role in
preference modeling. While semantic SIDs capture item content, their
unordered token structure lacks item-level cohesion and behavioral
transition information. The graph prompt addresses this limitation by
forming unified item representations, propagating transition signals,
and summarizing recent interactions into a behavior-aware user prompt.

Removing SID refinement also consistently degrades performance,
consistent with the SID quality improvements in
Table~\ref{tab:sid_quality}. Refinement removes redundant tokens,
preserves semantics, and reduces collisions, yielding more compact and discriminative supervision. The ranking loss aligns aggregated token scores with the final recommendation objective, while the contrastive loss aligns user representations with graph-enhanced target items.
Overall, the components provide complementary gains, with the graph
prompt contributing the largest improvement.

\subsection{Hyperparameter Study}
\begin{figure}[t]
    \centering
    \begin{subfigure}[b]{0.23\textwidth}
        \centering
        \includegraphics[width=\textwidth]{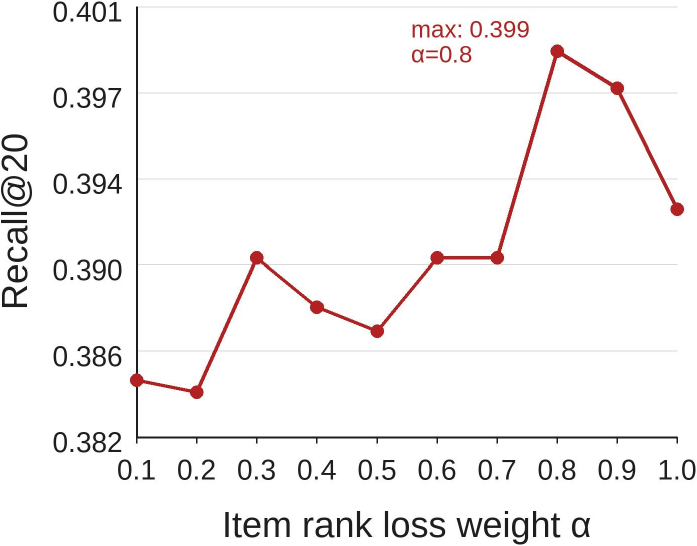}
        \caption{Luxury}
        \label{fig:hyperparameter_Luxury}
    \end{subfigure}
    % \hfill
    \begin{subfigure}[b]{0.23\textwidth}
        \centering
        \includegraphics[width=\textwidth]{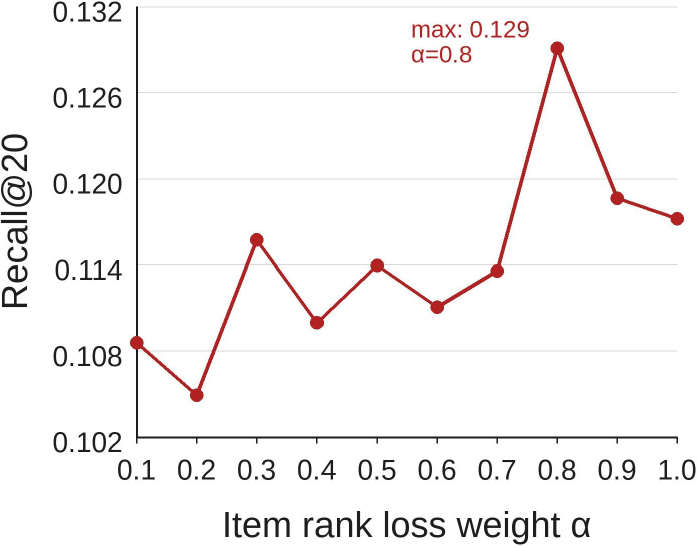}
        \caption{Scientific}
        \label{fig:hyperparameter_Scientific}
    \end{subfigure}
    % \hfill
    \begin{subfigure}[b]{0.23\textwidth}
        \centering
        \includegraphics[width=\textwidth]{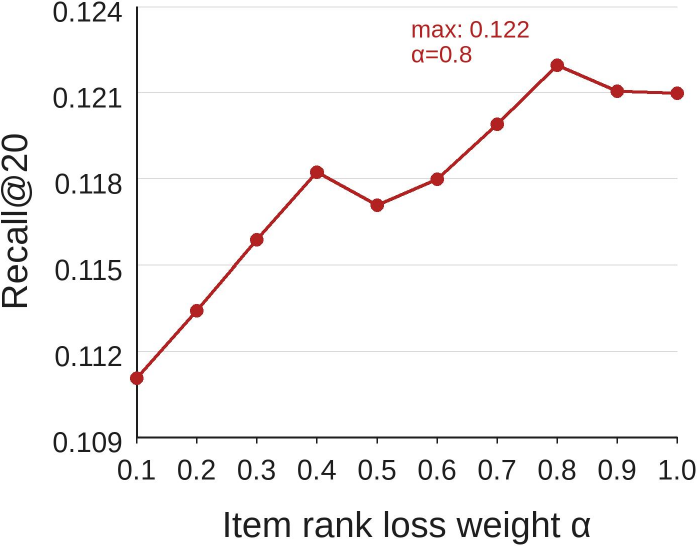}
        \caption{Instruments}
        \label{fig:hyperparameter_Instruments}
    \end{subfigure}
    \begin{subfigure}[b]{0.23\textwidth}
        \centering
        \includegraphics[width=\textwidth]{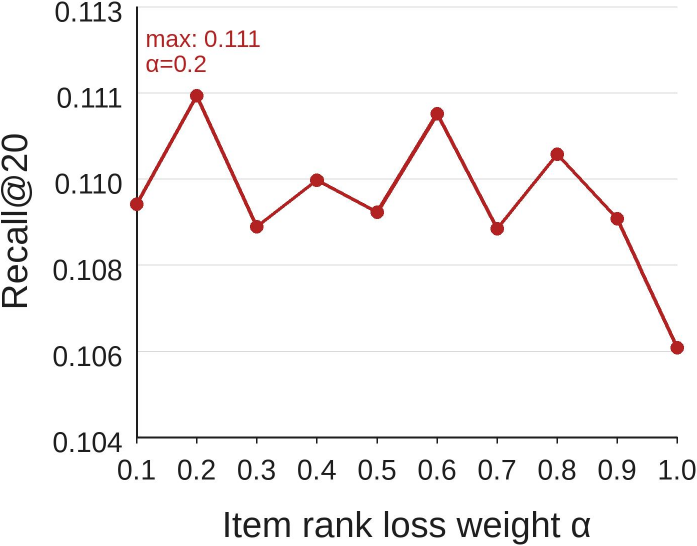}
        \caption{Arts}
        \label{fig:hyperparameter_Arts}
    \end{subfigure}
    \caption{Sensitivity to the ranking-loss weight $\alpha$.}
    \label{fig:hyperparameter}
    % \vspace{-10pt}
\end{figure}

\label{sec:hyperparameter}

We investigate the sensitivity of the item-level ranking loss weight $\alpha$, which balances token-level SID prediction and direct item ranking. We vary $\alpha$ from 0.1 to 1.0 while fixing all other hyperparameters.

As shown in Figure~\ref{fig:hyperparameter}, the preferred ranking weight varies across datasets. Luxury achieves its best result at $\alpha=0.8$, indicating that relatively dense interactions support stronger item-level ranking supervision. Although Scientific exhibits larger fluctuations, its optimum also occurs at $\alpha=0.8$. Recall@20 on Instruments increases steadily and peaks at $0.8$, which is consistent with the ranking objective helping distinguish items with similar predicted token sets. In contrast, Arts performs best at $\alpha=0.2$ and deteriorates at larger values, suggesting that excessive ranking supervision can overshadow token-level semantic learning in highly sparse and diverse item spaces. Overall, $\alpha=0.8$ provides the best trade-off on three datasets, while Arts benefits from a greater emphasis on token-level prediction.

\subsection{Efficiency Analysis}
\label{sec:efficiency}

\begin{table}[t]
\centering
\caption{Average online inference time per test instance on the Luxury item-cold dataset. $L$ and $B$ denote the SID length and beam size, respectively. The LLM computation column measures sequential decoding steps; offline SID construction is excluded.}
\label{tab:efficiency}
\resizebox{\linewidth}{!}{
\begin{tabular}{lccccc}
\toprule
\textbf{Method} & \textbf{Inference Mechanism}
& \textbf{LLM Computation}
& \textbf{Beam Size} & \textbf{Time (s)}
& \textbf{Speedup} \\
\midrule
LC-Rec
& Autoregressive decoding
& $\mathcal{O}(BL)$ decoding steps
& 100 & 6.824 & 1.00$\times$ \\
VaLiDRec
& Parallel token scoring
& One LLM forward pass
& -- & 0.078 & 87.49$\times$ \\
\bottomrule
\end{tabular}
}
% \vspace{-10pt}
\end{table}

We compare VaLiDRec with LC-Rec because both methods fine-tune the
same LLM backbone for SID-based recommendation, while adopting
different inference mechanisms. LC-Rec autoregressively generates
quantized SID tokens and applies beam search to obtain recommendation
candidates. Let $L$ denote the SID length and $B$ the beam size. Its
inference requires $\mathcal{O}(BL)$ sequential LLM decoding steps,
since each beam must be expanded over multiple generation steps.
In contrast, VaLiDRec predicts the scores of all semantic SID tokens
in parallel using a single LLM forward pass. The resulting token scores
are then aggregated over item SIDs to obtain item-level relevance
scores, without autoregressive generation or beam search. Therefore,
the number of LLM forward passes is constant with respect to $L$ and
$B$. This does not imply constant end-to-end complexity: full-catalog
item scoring requires
$\mathcal{O}\!\left(\sum_{j\in\mathcal{I}} |S_j|\right)$
lightweight token-to-item aggregation operations. SID construction is
performed offline and is excluded from online inference.

As reported in Table~\ref{tab:efficiency}, under the same LLM backbone,
hardware, and batch-size setting, LC-Rec requires an average of
6.824 seconds per test instance with a beam size of 100, whereas
VaLiDRec requires 0.078 seconds. This corresponds to an
$87.49\times$ speedup, demonstrating that replacing sequential SID
generation with one-pass token scoring substantially reduces the
online LLM inference overhead.

\subsection{SID Quality Analysis}
\label{sec:sid_quality}

To evaluate SID quality, we compare the Initial SID used in the
\textit{w/o SID Refinement} variant with the Refined SID produced by the complete Stage~1 process. The Initial SID contains
importance-selected candidates before pruning and collision handling. We assess semantic preservation, compactness, and item-level uniqueness.

\begin{table}[t]
\centering
\caption{Comparison of the semantic quality, compactness, and collision rates of the initial and refined SIDs.}
\label{tab:sid_quality}
\resizebox{\linewidth}{!}{
\begin{tabular}{llccccc}
\toprule
\textbf{Dataset} & \textbf{SID}
& \textbf{Quality}
& \textbf{$\geq$0.95}
& \textbf{Length}
& \textbf{Compression}
& \textbf{Collision$^\dagger$} \\
\midrule

\multirow{2}{*}{Luxury}
& Initial & 0.9023 & 13.66\% & 14.81 & 0.1493 & 1.90\% \\
& Refined & \textbf{0.9461} & \textbf{83.02\%}
& \textbf{8.81} & \textbf{0.1034} & \textbf{1.78\%} \\
\midrule

\multirow{2}{*}{Scientific}
& Initial & 0.9103 & 15.04\% & 15.14 & 0.1511 & 1.36\% \\
& Refined & \textbf{0.9495} & \textbf{84.12\%}
& \textbf{8.88} & \textbf{0.0982} & \textbf{0.77\%} \\
\midrule

\multirow{2}{*}{Instruments}
& Initial & 0.9214 & 26.86\% & 14.75 & 0.1980 & 0.83\% \\
& Refined & \textbf{0.9526} & \textbf{90.51\%}
& \textbf{8.27} & \textbf{0.1356} & \textbf{0.56\%} \\
\midrule

\multirow{2}{*}{Arts}
& Initial & 0.9245 & 26.00\% & 15.40 & 0.1890 & 0.97\% \\
& Refined & \textbf{0.9529} & \textbf{94.39\%}
& \textbf{7.56} & \textbf{0.1036} & \textbf{0.72\%} \\
\bottomrule
\end{tabular}
}
\vspace{1mm}
\begin{minipage}{\linewidth}
\footnotesize
$^\dagger$ Collision is measured before suffix-based disambiguation; the final collision ratio is 0 for all datasets.
\end{minipage}
% \vspace{-10pt}
\end{table}

As shown in Table~\ref{tab:sid_quality}, SID refinement consistently improves semantic quality while producing more compact identifiers. Average semantic quality increases from 0.9023--0.9245 to 0.9461--0.9529, while the proportion of SIDs above 0.95 rises from 13.66\%--26.86\% to 83.02\%--94.39\%. Meanwhile, the average SID length is reduced from about 15 tokens to 7.56--8.88, indicating that semantic-quality-aware pruning effectively removes redundant tokens while preserving item semantics. Collision-aware expansion further reduces semantic collisions, and suffix-based disambiguation eliminates all remaining item-level collisions. Overall, the refined SIDs are more semantically faithful, compact, and uniquely identifiable.

\section{Conclusion}

We presented VaLiDRec, a generative recommendation framework that constructs variable-length SIDs directly from the native vocabulary of pretrained LLMs. VaLiDRec selects informative metadata tokens through importance scoring and semantic-quality-aware pruning, resolves residual identifier collisions, and incorporates item-level structural relations and behavioral transitions through graph-aware soft prompts. By formulating recommendation as token-set prediction and token-level item scoring, it avoids autoregressive SID generation and beam search. Experiments on four datasets demonstrate consistent improvements over strong sequential and generative baselines, while the cold-start and efficiency results further show its effectiveness for unseen items and efficient online inference. In future work, we will extend VaLiDRec to multimodal item representations and investigate its transferability across recommendation domains.

%%
%% The acknowledgments section is defined using the ``acks'' environment
%% (and NOT an unnumbered section). This ensures the proper
%% identification of the section in the article metadata, and the
%% consistent spelling of the heading.
\begin{acks}
The Australian Research Council supports this work under the streams of Future Fellowship (Grant No. FT210100624), the Discovery Project (Grant No. DP240101108), and the Linkage Project (Grant No. LP230200892).
\end{acks}

%%
%% The next two lines define the bibliography style to be used, and
%% the bibliography file.
\bibliographystyle{ACM-Reference-Format}
\bibliography{mybib}

%%
%% If your work has an appendix, this is the place to put it.
\appendix

\end{document}